\documentclass[a4paper,twocolumn,fleqn]{article}

\usepackage{txfonts}               
\usepackage{helvet}
\usepackage{bm}                  
\usepackage{cite}                  
\usepackage{jpfr}                  

\usepackage{graphicx}

\usepackage{hyperref}

 \newcommand{\iotabar}{\mbox{$\,\iota\!\!$-}}
 \newcommand{\Fig}[1]{Fig.~\ref{fig:#1}}
 
 \newcommand{\Eqn}[1]{Eq.\,(\ref{eq:#1})}
 \newcommand{\Sec}[1]{Sec.\,\ref{sec:#1}}

 \renewcommand{\vec}[1]{\mbox{$\bf #1$}}
 
 \newcommand{\grad}{\mbox{\boldmath\(\nabla\)}}
 
 \newcommand{\curl}{  \mbox{\boldmath\(\nabla\times\)} }
 \newcommand{\dotv}{  \mbox{\boldmath\(\cdot\)} }
 \newcommand{\cross}{  \mbox{\boldmath\(\times\)} }
 \newcommand{\esub}[1]{ {\bf e}_{#1} }
 
 \newcommand{\dsub}[1]{ \partial_{#1} }
 \newcommand{\Proj}{\textsf{\textbf{P}}} 
 \newcommand{\be}{\begin{eqnarray}}
 \newcommand{\ee}{\end{eqnarray}}
 
 \newcommand{\Poincare}{Poincar{\'e}}

\begin{document}

\title{Unified Theory of Ghost and Quadratic-Flux-Minimizing Surfaces}


\author{Robert~L.~DEWAR, Stuart~R.~HUDSON\sup{1} and Ashley~M.~GIBSON}

\affiliation{Plasma Research Laboratories, Research School of Physics \& Engineering, The Australian National University, Canberra ACT 0200, Australia \\
  \sup{1}Princeton Plasma Physics Laboratory, PO Box 451, Princeton NJ 08543, USA \\
 }

\date{\today}

\email{robert.dewar@anu.edu.au}

\begin{abstract}
A generalized Hamiltonian definition of ghost surfaces (surfaces defined by an action-gradient flow) is given and specialized to the usual Lagrangian definition. Numerical calculations show uncorrected quadratic-flux-minimizing (QFMin) and Lagrangian ghost surfaces give very similar results for a chaotic magnetic field weakly perturbed from an integrable case in action-angle coordinates, described by $L = L_0 + \epsilon L_1$, where $L_0(\dot{\theta\,})$ (with $\dot{\theta\,}$ denoting $d\theta/d\zeta$) is an integrable field-line Lagrangian and $\epsilon$ is a perturbation parameter. This is explained using a perturbative construction of the auxiliary poloidal angle $\Theta$ that corrects QFMin surfaces so they are also ghost surfaces. The difference between the corrected and uncorrected surfaces is $O(\epsilon^2)$, explaining the observed smallness of this difference. An alternative definition of ghost surfaces is also introduced, based on an action-gradient flow in $\Theta$, which appears to have superior properties when unified with QFMin surfaces.
 \end{abstract}


\maketitle  


\section{\label{sec:intro}Introduction}
Recent calculations \cite{Hudson_Breslau_08} of heat diffusion along chaotic field lines show that the isotherms correspond very closely with the ``approximate'' magnetic surfaces, associated with magnetic island chains, known as ghost surfaces \cite{Hudson_Dewar_96}. These surfaces include the ``X-point'' and ``O-point'' closed field lines of their associated islands. (By ``O-point''  field line we mean either the elliptically stable field line at the center of an island or its hyperbolically unstable continuation if it has undergone a period-doubling bifurcation.)
Closed field lines extremize the magnetic action, $\oint\vec{A}\dotv\vec{dl}$, the hyperbolic X-point field lines in the chaotic separatrices being minima and the O-point field lines being minimax or saddle points of the action. Ghost surfaces are constructed by interpolating smoothly between these two closed-field-line classes by evolving the O-point field lines into the X-point field lines along paths of steepest descent of action, thus generating a family of ``pseudo-orbits,'' i.e. paths that come close to extremizing action.

An alternative approach to defining approximate magnetic surfaces passing through magnetic islands, is to use the quadratic-flux-minimizing (QFMin) surfaces introduced by Dewar, Hudson and Price \cite{Hudson_Dewar_96,Dewar_Hudson_Price_94}.
Ghost surfaces have nice mathematical properties but are difficult to construct and have no obvious physical interpretation. QFMin surfaces, on the other hand, have the computational attraction of being easy to construct, and the physical attraction of being defined in terms of a measure of the magnetic flux transport through the surface, but have been found to exhibit undesirable distortions in some circumstances.

However the definition of QFMin surfaces is not unique, as it depends on the choice of an auxiliary poloidal angle $\Theta$; which raises the question: Can $\Theta$ be chosen so that QFMin surfaces coincide with ghost  surfaces? This question has recently \cite{Hudson_Dewar_09} been answered in the affirmative, provided we modify the definition of ghost surface slightly, there being some freedom also in the definition of action-gradient flow. This raises the further question, accepting that ghost surfaces need to be redefined to achieve unification with QFMin surfaces, is the definition used in Ref.~\citen{Hudson_Dewar_09} optimal, or is there a still better class of unified ghost and QFMin surfaces?

In Secs.~\ref{sec:coords_flux}--\ref{sec:Lagrangian} we review the QFMin and ghost surface concepts, and in Sec.~\ref{sec:QFMinGhost} summarize the unification proposed in Ref.~\citen{Hudson_Dewar_09}. In Sec.~\ref{sec:QFMinGhostPert} we summarize its perturbative implementation in a test problem.

In Sec.~\ref{sec:NewGhost} we propose a new definition of ghost surfaces based on an action-gradient flow in the auxiliary angle $\Theta$, rather than the flow in the unperturbed canonical coordinate $\theta$ used in Ref.~\citen{Hudson_Dewar_09}. The new definition is not only simpler than that in Ref.~\citen{Hudson_Dewar_09}, in that it does not require a variable rate of flow along pseudo field lines, it also leads naturally to straight pseudo-field-line orbits in the $\Theta$-$\zeta$ plane. This is illustrated perturbatively for the same case used in Sec.~\ref{sec:QFMinGhostPert}.

\begin{figure}[htbp]
	\centering\includegraphics[width=5cm]{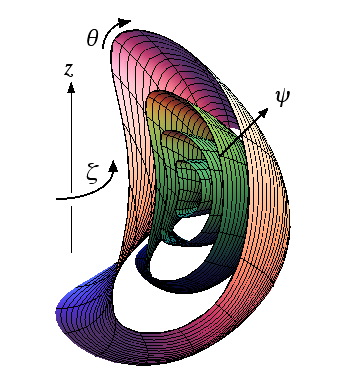}
	\caption{A sketch of the general curvilinear toroidal coordinate system described in the text.}
	\label{fig:Coordinates}
\end{figure}

\section{\label{sec:coords_flux}Coordinates and Fluxes}

We use a general curvilinear coordinate system $\psi,\theta,\zeta$, as shown in Fig.~\ref{fig:Coordinates}, where $\theta$ and $\zeta$ are poloidal and toroidal angles, respectively, and $\psi$ is a label for the coordinate surfaces that has the dimension of magnetic flux and is such that $\vec{B} = \grad\psi\cross\grad\theta + \grad\zeta\cross\grad\chi$.\footnote{Note that $\chi = \chi(\psi,\theta,\zeta)$ cannot in general be made a surface function in three-dimensional systems due to the generic nonintegrability of the magnetic field-line flow.}

The standard linear flux $\int\! dS\,\vec{n}\dotv\vec{B}$ through surface $\Gamma$: $\psi = \psi_{\Gamma}(\theta,\zeta)$, where $dS$ is an element of surface area and $\vec{n}$ is the unit normal,  
is independent of choice of coordinates and vanishes identically. Thus it is independent of the choice of $\Gamma$ also, whether it be a magnetic surface (invariant torus of the field-line flow) or otherwise.

\begin{figure}[htbp]
	\centering\includegraphics[width=6cm]{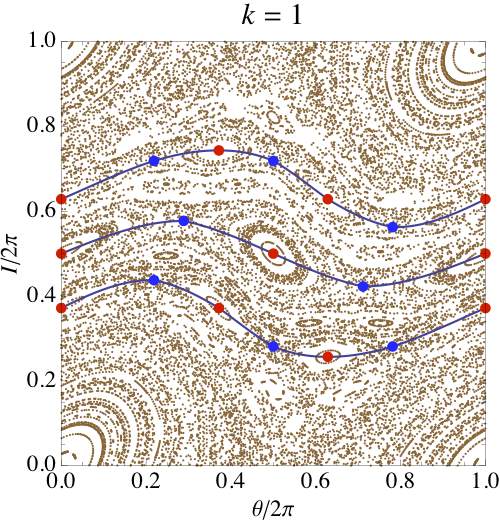}
	\caption{Poincar\'{e} plot for the Standard Map with the $p,q = 1,3$, 1,2 and 2,3 island chains, with associated action-minimizing (blue dots) and -minimax (red dots) periodic orbits. Also shown are almost-invariant curves interpolating between them.}
	\label{fig:StdMap}
\end{figure}

To measure the amount by which $\Gamma$ departs from being a magnetic surface, we are instead led to define the positive definite \emph{quadratic flux} \cite{Dewar_Hudson_Price_94}
\begin{equation}
	\varphi_2[\Gamma] \equiv \frac{1}{2}\int_0^{2\pi}\!\!\!\int_0^{2\pi}\!\!\!  d\theta d\zeta\,
	\frac{\vec{n}\dotv\vec{B}}{\vec{n}\dotv\grad\theta\cross\grad\zeta}
	\frac{\vec{n}\dotv\vec{B}}{\vec{n}\dotv\grad\Theta\cross\grad\zeta} \;,
	\label{eq:QFdef}
\end{equation}
where $\Theta$ is an auxiliary poloidal angle, not necessarily the coordinate angle $\theta$. Setting $\delta\varphi_2 = 0$ for arbitrary variations $\psi_{\Gamma}$ we find the Euler--Lagrange equation for QFMin pseudo-orbits\cite{Dewar_Hudson_Price_94,Hudson_Dewar_09} is $\vec{B}_{\nu}\dotv\grad\nu = 0$, where $\nu \equiv \vec{n}\dotv\vec{B}/\vec{n}\dotv\grad\Theta\cross\grad\zeta$ and the ``pseudo magnetic field'' $\vec{B}_{\nu}$ is defined by
\be
	\vec{B}_{\nu} \equiv \vec{B} - \nu \grad \Theta \cross \grad \zeta. \label{eq:pseudofield}
\ee
Thus $\nu$ parametrizes a continuous family of pseudo field lines, which can be thought of as \emph{orbits} of a dynamical system. We define a $p,q$-periodic pseudo-orbit ($p$ and $q$ being integers) as a path that is a closed loop, with $\theta$ increasing by $2\pi p$ when $\zeta$ increases by $2\pi q$, so the average rate of increase of $\theta$ along the path is the rational \emph{rotational transform} $\iotabar_{p,q} = p/q$.

For a given $\nu$ there are generally two distinct $p,q$-periodic orbits. In particular, when $\nu = 0$ we find the true periodic orbits associated with the X- and O-points of a $p,q$ island chain.

The family of $p,q$-periodic QFMin pseudo-orbits parametrized by $\nu$ defines an ``almost invariant'' surface with rotational transform $\iotabar_{p,q}$, which passes through the $p,q$ island chain as depicted in Fig.~\ref{fig:StdMap}.

\section{\label{sec:action_ghosts}Action and ghost surfaces}

Using the vector potential representation  $\vec{B} = \curl\vec{A}$, the magnetic action of a closed $p,q$-periodic path $\cal C$ is
\be {\cal S}[{\cal C}] \equiv \int_{\cal C} {\bf A}\dotv\vec{dl} \equiv \int_0^{2\pi q} {\bf A}\dotv\dot{\vec{r}}\,d\zeta, \label{eq:magneticaction}
\ee
where $\dot{\vec{r}}\equiv d\vec{r}/d\zeta$ along the path. Hamilton's Principle, $\delta {\cal S} = 0$, gives the Euler--Lagrange equation $\delta {\cal S}/\delta\vec{r} = \dot{\vec{r}} \cross\vec{B} = 0$, which implies $\dot{\vec{r}}$ is parallel to $\vec{B}$ as required.

Closed paths making $\cal S$ stationary are isolated, but the \Poincare--Birkhoff theorem implies there are at least two associated with each island chain, as illustrated in Fig.~\ref{fig:StdMap}, with the X-point orbits being minima of $\cal S$, while the O-point orbits are minimax (saddle) points of $\cal S$. The ghost-curve strategy for ``joining the dots'' is to generate a family of pseudo-orbits, labelled by a continuous parameter $\tau$, by flowing down the action gradient from minimax orbits to minimizing orbits,
\be \frac{D\vec{r}}{D\tau} = -\frac{\delta\cal S}{\delta\vec{r}}\dotv\Proj_{\rm ghost},\label{eq:gradflow}
\ee
where $\Proj_{\rm ghost}$ is a symmetric nonnegative dyadic whose choice is discussed below.
The union of this set of pseudo-orbits defines a \emph{ghost surface} (a \emph{ghost curve} being the Poincar\'e section of a ghost surface).

\begin{figure}[htbp]
	\centering\includegraphics [width=6cm]{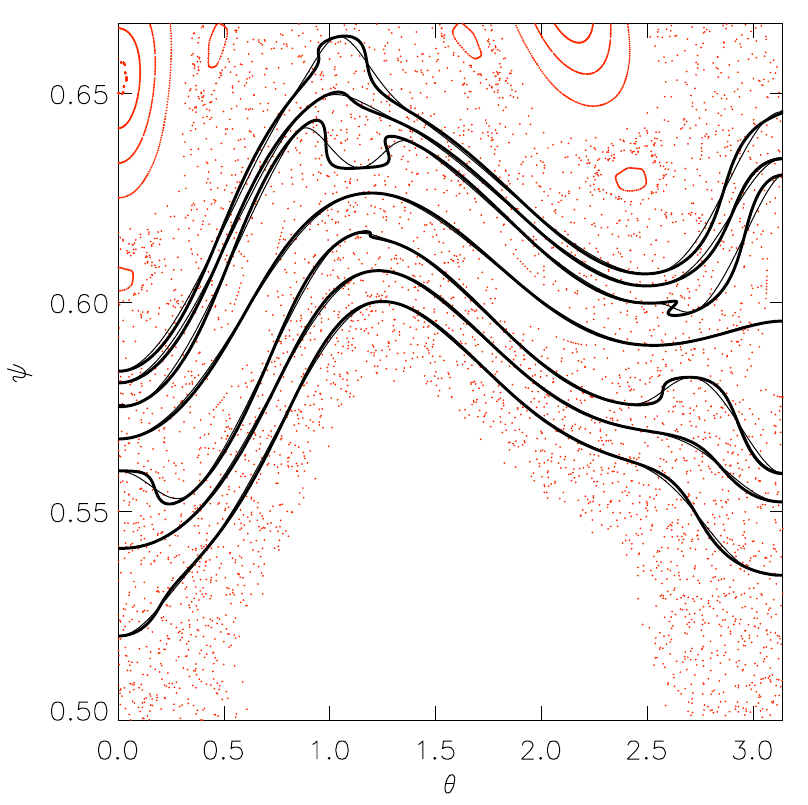}
	\caption{A comparison of uncorrected ($\Theta = \theta$) QFMin curves (thick lines) and ghost curves (thin lines) for the more strongly chaotic case described in Ref.~\citen{Hudson_Dewar_09}. Some cases where QFMin curves violate the graph property are seen.}
	\label{fig:comparisonhigh}
\end{figure}

Ghost curves and \emph{uncorrected} QFMin curves (i.e. curves for which the auxiliary angle $\Theta$ has been taken to be the same as the coordinate angle $\theta$) are compared in Fig.~\ref{fig:comparisonhigh}. For the lower-order $p,q$ cases the two definitions of almost-invariant curves are almost indistinguishable. However, higher-order uncorrected QFMin curves can cease to be graphs over $\theta$ and do not agree with the isotherms found by Hudson and Breslau \cite{Hudson_Breslau_08}. Thus we are led to seek a way to choose $\Theta$ so that the two approaches may be reconciled, keeping the physical appeal and ease of implementation of the QFMin approach while retaining the superior mathematical properties of ghost surfaces.

\section{\label{sec:Lagrangian}Lagrangian approach}

We first map this problem onto one of Lagrangian dynamics in order to make better contact with the mathematical literature, e.g. Ref.~\citen{Gole_01}.

The vector potential $\vec{A}=\psi\grad\theta-\chi(\psi,\theta,\zeta)\grad\zeta$ gives our assumed form for $\vec{B}$ and leads to the action integral ${\cal S}[{\cal C}] =\oint (\psi d\theta - \chi d\zeta)$, the first variation being
\be
	\delta {\cal S}  =  \int_0^{2\pi q}\!
\left[
     \left(\dot{\theta\,} - \dsub{\psi}\chi \right)\delta\psi
   - \left(\dot{\psi} + \dsub{\theta}\chi\right)\delta\theta
\right]  d\zeta \;.
\ee
Setting $\delta{\cal S} = 0$ for all $\delta\psi$ and $\delta\theta$ leads to Hamiltonian equations of motion with $\theta$ as generalized coordinate, $\psi$ as canonical momentum, $\chi$ as Hamiltonian, and $\zeta$ as time: $\dot{\theta\,}  =  \dsub{\psi}\chi$ and
$\dot{\psi}  =  -\dsub{\theta}\chi$.

The transition to the Lagrangian approach is made in the usual way, by solving $\dot{\theta\,}  =  \dsub{\psi}\chi$ to give $\psi$ as a function of $\dot{\theta\,}$ and writing the action as ${\cal S} = \int \!\!L\, d\zeta$, where $L(\theta,\dot{\theta\,},\zeta) \equiv \psi(\theta,\dot{\theta\,},\zeta)\,\dot{\theta\,} - \chi(\psi,\theta,\zeta)$. The first variation now becomes $\delta {\cal S} = \int_0^{2\pi q}\!\delta\theta\,(\delta {\cal S}/\delta\theta)\,d\zeta $, the functional derivative (action gradient) being
\be
	\frac{\delta{\cal S}}{\delta\theta} = \frac{\partial L}{\partial\theta}
	- \frac{d}{d\zeta}\frac{\partial L}{\partial\dot{\theta\,}} \;.
\ee

Specialization of the general ghost-orbit gradient flow \Eqn{gradflow}  to the Lagrangian form is effected by taking $\Proj_{\rm ghost} = \epsilon^{-1}\esub{\psi}\esub{\psi} + \mu^{-1}\esub{\theta}\esub{\theta}$ and taking the limit $\epsilon \to 0$ to enforce the constraint $\dot{\theta\,}  =  \dsub{\psi}\chi$. Then the gradient flow becomes
\be	\frac{D\theta}{D\tau} & = & -\frac{1}{\mu(\zeta)}\frac{\delta{\cal S}}{\delta\theta}
	\;, \label{eq:Laggradflow}
\ee
where $\mu(\zeta) = O(1)$ allows us to generalize the form assumed in Ref.~\citen{Hudson_Dewar_96} slightly, which we shall find necessary for reconciliation to be possible.

\section{Reconciling QFMin and Lagrangian Ghost surfaces}\label{sec:QFMinGhost}

We now seek to choose $\Theta$ so that QFMin pseudo-orbits are also the Lagrangian ghost pseudo-orbits defined above.
We require \cite{Hudson_Dewar_09} $\partial_{\psi}\Theta = 0$.
 That is,
\be \Theta \equiv \Theta(\theta,\zeta) \;.
	\label{eq:thetaTransfn}
\ee
Then $\Psi_{\nu} = \Psi$, $L_{\nu} = L - \nu\Theta$, $\partial L_{\nu}/\partial\dot{\theta\,} = \partial L/\partial\dot{\theta\,} $, and members of our family of QFMin pseudo-orbits
\be \theta = \theta(\zeta|\Theta_0) \;,
	\label{eq:thetafamily}
\ee
where $\Theta_0$ is, as yet, an arbitrary label, satisfy the Euler--Lagrange equation
\be \frac{\delta{\cal S}_{\nu}}{\delta\theta}  = 
	\frac{\delta{\cal S}}{\delta\theta} - \nu(\Theta_0)\Theta_{\theta}(\theta,\zeta) = 0\;, 
	\label{eq:LagVar1recon}
\ee
with $\Theta_{\theta}(\theta,\zeta) \equiv \partial\Theta(\theta,\zeta)/\partial\theta$.

To reconcile  QFMin and ghost orbits we require that the family of pseudo-orbits defined by Eqs~(\ref{eq:thetafamily}) and (\ref{eq:LagVar1recon}) is the same family as is generated by \Eqn{Laggradflow}. Thus the labels $\Theta_0$ and $\tau$ must be functionally dependent: $\tau = \tau(\Theta_0)$, $d\tau = \tau'(\Theta_0)d\Theta_0$. Eliminating $\delta{\cal S}/\delta\theta$ between \Eqn{LagVar1recon} and \Eqn{Laggradflow} and observing that $\Theta_{\theta} \equiv (D\Theta/D\Theta_0)/(D\theta/D\Theta_0)$, where $D\theta/D\Theta_0 \equiv \partial\theta(\zeta|\Theta_0)/\partial\Theta_0$, we find the \emph{reconciliation condition}
\be \frac{D\Theta}{D\Theta_0}   = -\frac{\mu(\zeta)}{\tau'(\Theta_0)\nu(\Theta_0)}\left(\frac{D\theta}{D\Theta_0}\right)^2 \;.
	\label{eq:reconcilcon}
\ee
We now \emph{define} $\Theta_0$ so that, for all $\Theta_0$,
\be \tau'(\Theta_0)\nu(\Theta_0) & \equiv & -1 \;, \nonumber \\
	\theta(\zeta|\Theta_0+2\pi) & \equiv & \theta(\zeta|\Theta_0) + 2\pi \;,
\ee
choosing $\mu(\zeta)$ so that \Eqn{reconcilcon} satisfies the \emph{solvability condition} that the integral of both sides with respect to $\Theta_0$ over the interval $[0,2\pi]$ must be $2\pi$, giving
\be \mu(\zeta) = \left[\int_0^{2\pi}\frac{d\Theta_0}{2\pi}\left(\frac{\partial\theta(\zeta|\Theta_0)}{\partial\Theta_0}\right)^2\right]^{-1} \;.
\label{eq:solvability} \ee

\section{Perturbative construction of QFMin-ghost surfaces}\label{sec:QFMinGhostPert}

For example, consider
\be L = \frac{\dot{\theta\,}^2}{2} - \epsilon\!\!\!\!\sum^{\infty}_{m,n=-\infty}\!\!\!\! V_{m,n}\exp(im\theta-in\zeta) \;, \label{eq:modelL}
\ee
with the reality condition $V^{*}_{m,n} = V_{-m,-n}$, and $\epsilon$ the expansion parameter. (Defining the largest perturbation coefficient, $V_{2,1}$, to be 1, $\epsilon = 10^{-3}$ for the case shown in \Fig{comparisonhigh}.)

As the unperturbed system is integrable, the expansions of $\nu(\Theta_0)$, $\mu(\zeta)$, and $\Theta(\theta,\zeta)$ are of the form
\be \nu & = & \epsilon\nu_1 + \epsilon^2\nu_2 + \ldots \;, \nonumber\\
	\mu & = & \epsilon \mu_1 + \epsilon^2 \mu_2 + \ldots \;, \\
	\Theta & = & \theta + \sum_{m,n}\left(\epsilon \Theta^{(1)}_{m,n} 
	+ \ldots\right)\exp i(m\theta - n\zeta) \nonumber \;,
	 \label{eq:epsExpansion}
\ee
and the $(p,q)$ QFMin pseudo-orbits are expanded as
\be \theta(\zeta|\Theta_0) & = & \iotabar_{p,q}\zeta + \Theta_0 + \sum_{m,n}\left(\epsilon\theta^{(1)}_{m,n} + \epsilon^2\theta^{(2)}_{m,n}
+ \ldots\right)
	\nonumber\\
		& & \quad \times\exp i\left[\left(m\iotabar_{p,q}-n\right)\zeta + m\Theta_0\right] \;,
	 \label{eq:OrbitepsExpansion}
\ee
where $\iotabar_{p,q} \equiv p/q$.

At $O(\epsilon)$ we find $\mu_1 = 0$. Also, $\Theta^{(1)}_{m,n}$ is not used in the calculation of $\theta^{(1)}_{m,n}$. Thus, \emph{to first order, uncorrected ghost and QFMin pseudo-orbits are identical}. This explains why the low-order uncorrected ghost and QFMin orbits appear almost identical in \Fig{comparisonhigh}.

Explicit expressions for $\nu_1$, $\mu_2$, $\theta^{(1)}_{m,n},$ $\theta^{(2)}_{m,n}$ and $\Theta^{(1)}_{m,n}$ are given in Ref.~\citen{Hudson_Dewar_09}, but, as there is a misprint in $\theta^{(2)}_{m,n}$, we give below a corrected version,
\be \theta^{(2)}_{m,n} & = & \frac{i\bar{\delta}_{mp,nq}}{(m\iotabar_{p,q} - n)^2}\sum^{\phantom{xxxx}\prime}_{m',n'}
	\frac{N^{\theta}_{m,m',n,n'}}
		{(m'\iotabar_{p,q} - n')^2} \nonumber\\
	&& \quad\quad\quad\quad\quad\mbox{}\times V_{m+m',n+n'}V_{m',n'}^{*}
\;, \label{eq:theta2old}
\ee
where $N^{\theta}_{m,m',n,n'} \equiv m'(m+m')\left[m+m' + m'\delta_{(m+m')p,(n+n')q}\right]$. Also,
$\bar{\delta}_{mp,nq} \equiv 1-\delta_{mp,nq}$, where $\delta_{mp,nq} \equiv 1$ when $mp = nq$, is 0 otherwise, and the prime on the sum over $m'$ and $n'$ indicates that the resonant terms, $m'p = n'q$, are to be deleted.

\section{$\Theta$ ghost surface formulation}\label{sec:NewGhost}

In Sec.~\ref{sec:Lagrangian} we defined ghost surfaces in terms of an action-gradient flow in $\theta$, but defined QFMin surfaces in terms of $\Theta$. Here we further unify the formulation by defining the action-gradient flow also in terms of $\Theta$,
\be	\frac{D\Theta}{D T}  =  -\frac{\delta{\cal S}}{\delta\Theta} 
	\equiv -\vartheta_{\Theta}\frac{\delta{\cal S}}{\delta\theta}
	\;, \label{eq:ThetaLaggradflow}
\ee
where $T$ is the timelike label for $\Theta$ ghost orbits [so $dT = T'(\Theta_0)d\Theta_0$] and the inverse function $\vartheta(\Theta,\zeta)$ is defined by solving \Eqn{thetaTransfn} for $\theta$ as a function of $\Theta$. 

From \Eqn{LagVar1recon} we also have, for corrected QFMin orbits,
\be \frac{\delta{\cal S}}{\delta\Theta} = \nu(\Theta_0)
	\;. \label{eq:LagVar2recon}
\ee
Eliminating $\delta{\cal S}/\delta\Theta$ between \Eqn{ThetaLaggradflow} and \Eqn{LagVar2recon} we find the new reconciliation condition
\be \frac{D\Theta}{D\Theta_0}   = -T'(\Theta_0)\nu(\Theta_0) = 1 \;,
	\label{eq:reconcilcon2}
\ee
where, in the second equality, we have defined $\Theta_0$ so that $T'(\Theta_0)\nu(\Theta_0) \equiv -1$, which satisfies the solvability condition that $\int_0^{2\pi}d\Theta_0\,D\Theta/D\Theta_0 = 2\pi$ without the need to introduce a factor like the $\mu(\zeta)$ required in \Sec{QFMinGhost}.

Furthermore, \Eqn{reconcilcon2} is satisfied by choosing the $\zeta$-dependence of the pseudo-orbit labeling  so that \be
	\Theta(\zeta|\Theta_0) = \Theta_0 + \iotabar\zeta \;,
	\label{eq:SFL}
\ee
conjugating the $\theta$ dynamics to straight-field-line dynamics (cf. Ref~\citen{Dewar_Meiss_92}). Thus the $\Theta$ ghost surface approach appears to be much superior to $\theta$ ghost surface approach, but needs to be tested in practice.

As a first test we adapt the perturbative calculation in \Sec{QFMinGhostPert} to  $\Theta$ ghost surfaces. First, expanding $\vartheta(\Theta,\zeta) = \Theta + \sum_{m,n} \theta_{m,n}\exp(im\Theta - in\zeta)$ and using \Eqn {SFL}, we recover \Eqn {OrbitepsExpansion}. This ansatz is to be inserted in \Eqn{LagVar2recon}, written in the form
\be \left(\frac{dL_{\dot{\theta\,}}}{d\zeta} - L_{\theta}\right)\vartheta_{\Theta} + \nu(\Theta_0) = 0
	\;. \label{eq:LagVar2explicit}
\ee
From \Eqn{modelL}, $dL_{\dot{\theta\,}}/d\zeta = -\sum_{m,n} (m\iotabar - n)^2\theta_{m,n}\exp[im\Theta_0 + i(m \iotabar - n)\zeta]$, $L_{\theta} = -\epsilon\sum_{m,n} im V_{m,n}\exp [im\theta(\zeta|\Theta_0) - in\zeta]$, and $\vartheta_{\Theta} = 1+ \sum_{m,n} im\theta_{m,n}\exp[im\Theta_0 + i(m\iotabar - n)\zeta]$. Inserting these into \Eqn{LagVar2explicit}, and equating the LHS to zero at $O(\epsilon)$, we get
\be \nu_1(\Theta_0)  =  -\!\!\!\!\sum^{\infty}_{m,n=-\infty}\!\!\!\! im\delta_{mp,nq}V_{m,n}e^{im\Theta_0}\;, \label{eq:nu1}
\ee
and
\be \theta^{(1)}_{m,n}  =  \frac{im\bar{\delta}_{mp,nq}V_{m,n}}{(m\iotabar_{p,q} - n)^2}
\;, \label{eq:thetaTheta1}
\ee
which are the same results as in Ref.~\citen{Hudson_Dewar_09}, showing that $\theta$ and $\Theta$ ghost orbits (and, by definition, reconciled QFMin orbits) are equivalent at this order.

At $O(\epsilon^2)$ we find
\be \nu_2  & = &  -i\sum_{m,n}\delta_{mp,nq}\sum^{\phantom{xxxx}\prime}_{m',n'}
	\frac{N^{\Theta}_{m,m',n,n'}}
	{(m'\iotabar_{p,q} - n')^2}  \nonumber\\
	&& \quad\quad\quad\quad\quad\mbox{}\times V_{m+m',n+n'}V_{m',n'}^{*}e^{im\Theta_0}
\;, \label{eq:nu2}
\ee
and
\be \theta^{(2)}_{m,n} & = & \frac{i\bar{\delta}_{mp,nq}}{(m\iotabar_{p,q} - n)^2}\sum^{\phantom{xxxx}\prime}_{m',n'}
	\frac{N^{\Theta}_{m,m',n,n'}}
		{(m'\iotabar_{p,q} - n')^2} \nonumber\\
	&& \quad\quad\quad\quad\quad\mbox{}\times V_{m+m',n+n'}V_{m',n'}^{*}
\;, \label{eq:theta2new}
\ee
where $N^{\Theta}_{m,m',n,n'} \equiv m'(m+m')\left[m+m' - m'\delta_{(m+m')p,(n+n')q}\right]$
(the sign of the $\delta_{(m+m')p,(n+n')q}$ term in this expression being the opposite of that for $\theta$ ghost orbits).

\section{Conclusion}
We have reviewed the motivation and formulation of a recently published \cite{Hudson_Dewar_09} unification of ghost and quadratic-flux-minimizing surfaces, and have presented a new formulation that appears more natural and more useful in that it could be used to form a straight pseudo-field-line coordinate system.

More work remains to be done to develop a numerical algorithm for constructing such ghost/QFMin surfaces at arbitrary order of perturbation, and to show they preserve the desirable properties of our previous formulation of ghost surfaces.

The relation to heat transport in chaotic fields also needs to be investigated more deeply.

\bibliographystyle{prsty}
\bibliography{RLDBibDeskPapers}

\end{document}